\documentclass[twocolumn,prb,amsmath,amssymb,longbibliography,floatfix,superscriptaddress]{revtex4-2}
\usepackage{amsmath}
\usepackage{amssymb}
\usepackage{color}
\usepackage{physics}
\usepackage{graphicx}
\usepackage{tabularx}
\usepackage{dcolumn} % Align table columns on decimal point
\usepackage{bm}      % bold math
\usepackage{array}   % table making
\usepackage{amsfonts}
\usepackage{url}
\usepackage[bookmarks=false,colorlinks,citecolor=red]{hyperref}
\usepackage{natbib}
\usepackage{mathtools}
\usepackage{mwe}

\DeclareUnicodeCharacter{2212}{-}

\begin{document}

\include{MyCommand}
\newcommand{\AK}[1]{{\color{red}{#1}}}
\newcommand{\BR}[1]{{\color{blue}{#1}}}
\newcommand{\prateek}[1]{{\color{green}{#1}}}
\newcommand{\bkc}[1]{{\color{red}{#1}}}

%\title{Molecular dynamics on quantum hardware through machine learning}
\title{Quantum Hardware-Enabled Molecular Dynamics via Transfer Learning}
%\title{refining neural network potential energy surfaces with quantum data}
\author{Abid Khan}
\email{aakhan3@illinois.edu}
\affiliation{Department of Physics, University of Illinois Urbana-Champaign, Urbana, IL, United States 61801}
\affiliation{USRA Research Institute for Advanced Computer Science (RIACS), Mountain View, CA, 94043, USA}
\affiliation{NASA Ames Research Center, Moffett Field, CA, 94035, USA}

\author{Prateek Vaish}
%\email{prateek_vaish@brown.edu}
\affiliation{Department of Chemistry, Brown University, Providence, RI 02912}

\author{Yaoqi Pang}
%\email{yaoqi_pang@brown.edu}
\affiliation{Department of Chemistry, Brown University, Providence, RI 02912}

\author{Nikhil Kowshik}
%\email{yaoqi_pang@brown.edu}
\affiliation{Department of Chemistry, Brown University, Providence, RI 02912}

\author{Michael S. Chen}
%\email{mc10050@nyu.edu}
\affiliation{Department of Chemistry and Simons Center for Computational Physical Chemistry, New York University, New York, NY 10003}

\author{Clay H. Batton} 
\affiliation{Department of Chemistry, Stanford University, Stanford, CA 94305, USA}

\author{Grant M. Rotskoff}
\affiliation{Department of Chemistry, Stanford University, Stanford, CA 94305, USA}
\affiliation{Institute for Mathematical and Computational Engineering, Stanford University, Stanford, CA 94305, USA}

\author{J. Wayne Mullinax}%
\affiliation{KBR, Inc., Intelligent Systems Division, NASA Ames Research Center, Moffet Field, CA  94035, USA}

\author{Bryan  K. Clark}
%\email{bkclark@illinois.edu}
\affiliation{Department of Physics, University of Illinois Urbana-Champaign, Urbana, IL, United States 61801}
\affiliation{IQUIST and Institute for Condensed Matter Theory and NCSA Center for Artificial Intelligence Innovation, University of Illinois at Urbana-Champaign, IL 61801, USA}

\author{Brenda M. Rubenstein}
%\email{brenda_rubenstein@brown.edu}
\affiliation{Department of Chemistry, Brown University, Providence, RI 02912}
\affiliation{Department of Physics, Brown University, Providence, RI 02912}

\author{Norm M. Tubman}%
\email{norman.m.tubman@nasa.gov}
\affiliation{NASA Ames Research Center, Moffett Field, CA, 94035, USA}

\begin{abstract}
The ability to perform \textit{ab initio} molecular dynamics simulations using potential energies calculated on quantum computers would allow virtually exact dynamics for chemical and biochemical systems, with substantial impacts on the fields of catalysis and biophysics. However, noisy hardware, the costs of computing gradients, and the number of qubits required to simulate large systems present major challenges to realizing the potential of dynamical simulations using quantum hardware. Here, we demonstrate that some of these issues can be mitigated by recent advances in machine learning. By combining transfer learning with techniques for building machine-learned potential energy surfaces,  we propose a new path forward for molecular dynamics simulations on quantum hardware. We use transfer learning to reduce the number of energy evaluations that use quantum hardware by first training models on larger, less accurate classical datasets and then refining them on smaller, more accurate quantum datasets. We demonstrate this approach by training machine learning models to predict a molecule's potential energy using Behler-Parrinello neural networks. When successfully trained, the model enables energy gradient predictions necessary for dynamics simulations that cannot be readily obtained directly from quantum hardware. To reduce the quantum resources needed, the model is initially trained with data derived from low-cost techniques, such as Density Functional Theory, and subsequently refined with a smaller dataset obtained from %a Variational Quantum Eigensolver 
the optimization of the Unitary Coupled Cluster ansatz. We show that this approach significantly reduces the size of the quantum training dataset while capturing the high accuracies needed for quantum chemistry simulations. The success of this two-step training method opens up more opportunities to apply machine learning models to quantum data, representing a significant stride towards efficient quantum-classical hybrid computational models.
\end{abstract}
\date{\today}
 \maketitle

%As first enunciated by Feynman in his famous lecture on quantum computing~\cite{Feynman1982}, one of the most celebrated potential applications of quantum computing is to solve quantum chemistry problems outside the reach of conventional classical electronic structure algorithms~\cite{Cao_ChemRev,McArdle_RMP_2020,bauer_quantum_2020}.
Quantum computers are expected to have a transformational impact on quantum chemistry by extending the reach of classical electronic structure algorithms while also improving accuracy~\cite{Feynman1982, Cao_ChemRev,McArdle_RMP_2020,bauer_quantum_2020}.
Quantum chemistry problems, which include such technologically significant problems as predicting drug binding sites and designing novel catalysts, involve solving the Schrödinger equation for different atomic geometries to a sufficiently high level of accuracy \cite{friesner2005ab,szabo2012modern,helgaker2013molecular}. Many classical electronic structure methods such as Coupled Cluster (CC) Theory \cite{bartlett2007coupled,solomonik2014massively,doi:10.1021/ct200809m,matthews2020coupled,stein2014seniority}, selected Configuration Interaction (sCI) \cite{tubman2016deterministic,holmes2016heat,schriber2016communication,coe2018machine,coe2023analytic,mejuto2022effect,williams2023parallel,tubman2018efficient,tubman2020modern,garniron2017hybrid,garniron2018selected,lesko2019vibrational,pineda2021chembot}, Density Matrix Renormalization Group (DMRG)~\cite{olivares2015ab,white1992density,Schollw_ck_2011,Baiardi_2020}, and Quantum Monte Carlo (QMC) \cite{foulkes2001quantum,motta2018ab,kim2018qmcpack,tubman2011prospects,shulenburger2013quantum} methods have been developed to undertake these prediction tasks, often with sufficiently high accuracy, but also at a significant computational cost~\cite{eriksen2020ground}. When considering different possible simulation approaches, general-purpose exact methods scale exponentially with system size,  while highly accurate but approximate methods scale as a high-degree polynomial with system size.  Many important problems in chemistry and materials science are beyond the reach of what we can simulate classically with highly accurate methods~\cite{Reiher_2017}. %limiting their application to relatively small systems. 
%Moreover, many quantum chemistry methods struggle to describe systems with long-range correlations that cannot be addressed by carefully selecting active spaces or fragments.
Quantum computers, particularly once they reach fault tolerance, could solve challenging quantum chemistry problems in polynomial time, making problems previously too large or too correlated to treat with classical methods tractable. 

While we expect many quantum algorithms, like quantum phase estimation~\cite{nelson2024assessment,russo2021evaluating}, to require fault-tolerant quantum computers, methods already exist for using near-term quantum devices to solve problems in quantum chemistry~\cite{Lanyon2010,Tilly_2022,klymko2022real,kremenetski2021simulation,kremenetski2021quantum,robledomoreno2024chemistry,Huggins_2022,pathak2024requirements,rubin2023quantum,2021arXiv210914114B,gustafson2024surrogate,shen2023estimating,amsler2023quantumenhanced,sokolov2024quantumenhanced,nakaji2024generative}. Several recent papers have demonstrated the promise of the variational quantum eigensolver (VQE) for obtaining accurate electronic structure energies for applications in catalysis and chemical reactions. For example, a recent paper demonstrated on an 8-qubit IBM device that one can compute an accurate activation barrier of a Diels-Alder Reaction with VQE~\cite{liepuoniute2024simulation}. Additionally, quantum algorithms specific to catalysis have been presented~\cite{PhysRevResearch.3.033055} along with methods to sample rare conformational transitions for simulating thermal fluctuations in metastable states~\cite{Ghamari2022}. 

%Despite 
Recent efforts modeling molecular systems on quantum computers have made substantial progress in obtaining energies as the main quantity of interest. While energies help benchmark against spectroscopic data and determine ground states, other important data for understanding chemical processes include predicting molecular geometries \cite{schlegel2011geometry}, reaction pathways \cite{Henkelman_JCP,bowman2011high}, and dynamics \cite{tuckerman_ab_2002} - all of which necessitate knowledge of gradients \cite{iyer2024forcefree}. Within the Born-Oppenheimer approximation, gradients of a system's potential energy surface, such as forces, can be used to relax molecular geometries to their minima and solve Newton's equations of motion to predict dynamics. Nonetheless, recent literature suggests that calculating forces using nuclear gradients directly on quantum hardware is a costly undertaking~\cite{OBrienPRR,10.1063/5.0046930}. 

While methods that avoid calculating gradients ameliorate these issues~\cite{tubman2015molecular}, machine learning is an increasingly attractive option when quantum hardware is limited. Over the past decade, these various machine learning methods have been used to learn potential energy surfaces and their gradients for dynamics using electronic structure information from Density Functional Theory (DFT), dramatically accelerating conventional \textit{ab initio} molecular dynamics calculations. Examples of these machine learning methods include Behler-Parinello Neural Networks (BPNNs) \cite{PhysRevLett.98.146401,artrith2016implementation,artrith2021best,artrith2017efficient}, Gaussian Approximation Potentials \cite{GAPs_PRL,BokdamPRL}, Gradient-Domain Machine Learning \cite{chmiela2017machine}, deep potentials \cite{zhang2018deep} and graph neural networks \cite{batatia2022mace,batzner20223}. 

\begin{figure*}
    \centering
    \includegraphics[width=\linewidth]{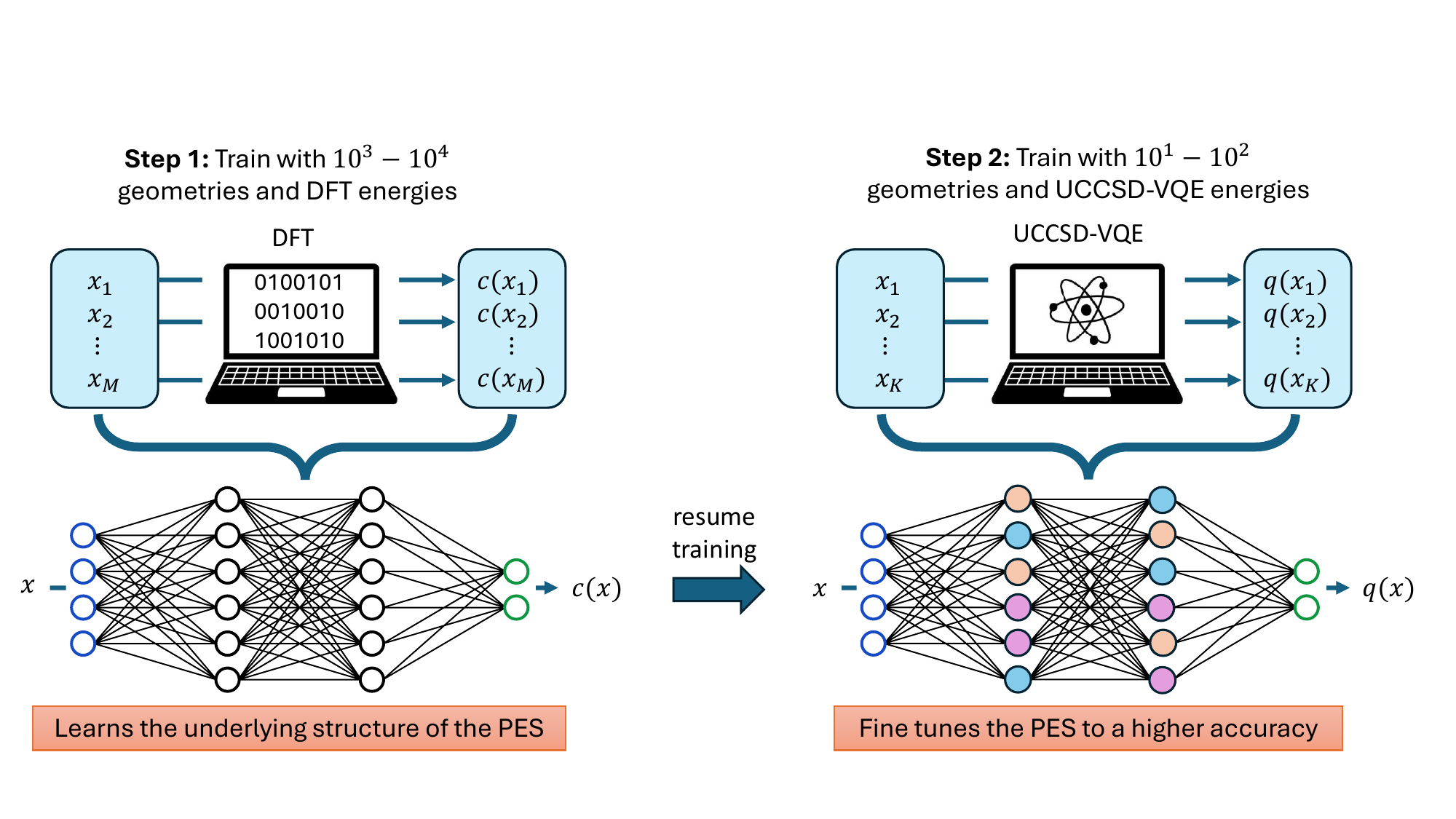}
    \caption{Overview of the transfer learning procedure. \textbf{(Step 1)}: A potential energy surface, represented by a neural network, is trained with a large dataset containing geometries and their respective DFT energies. \textbf{(Step2)}:  After completing the DFT training, we then resume training the same network with a smaller dataset containing geometries and their respective VQE energies.}
    \label{fig:fig1}
\end{figure*}

More recently, these ML methods have been enhanced to further correct DFT potential energy surfaces with information from high-accuracy electronic structure methods, including coupled cluster theories \cite{schran2019automated,daru2022coupled,bowman2022delta,nandi2021delta,qu2021breaking,Chen2022} and quantum Monte Carlo methods \cite{Cancan2022,Chen2022,niu2023stable,archibald2018gaussian,ryczko2022machine}, which can face scaling and other challenges when computing energy gradients \cite{Cancan2022,ceperley2024training}. A major challenge in using these high-cost, high-accuracy electronic structure methods for ML is the large dataset required to train neural network potentials. To address this, transfer learning techniques have been developed to improve data efficiency by initially learning the features of a given chemical system using data from a low-cost, low-accuracy method and then refining the model with a substantially small dataset from the high-cost, high-accuracy method. While previous work involved transfer-learning between two classical methods~\cite{Chen2022}, in this work, we leverage transfer-learning between a classical method and one that is performed with quantum hardware as illustrated in  Fig.~\ref{fig:fig1}. To that end, we demonstrate an efficient molecular dynamics engine for determining nuclear forces and performing molecular dynamics simulations based on energies from quantum computers.

Our paper is organized as follows: we describe our methods, including how we generated our training data sets, our transfer, and active learning techniques, and our molecular dynamics simulations in Section \ref{sec:methods}. We then illustrate how our methods perform and quantify the accuracy of our energy predictions and molecular dynamics simulations in Section \ref{sec:res}. Lastly, we discuss how our basic framework can be improved and potential applications in Section \ref{sec:conc}.

\section{Methods \label{sec:methods}}

%Our approach for molecular dynamics simulations on quantum hardware involves a couple of key steps for learning forces from quantum data.  
%In particular our approach involves transfer learning -
To make efficient use of a small number of high-accuracy VQE energies, we employ transfer learning. We represent the potential energy surface as a neural network, which takes the coordinates of the system as an input and outputs an energy. Importantly, this neural network function can be evaluated and differentiated efficiently on classical computers. We initially train this neural network potential on relatively low-cost DFT calculations and subsequently update the parameters of our model using higher-cost but more accurate calculations \cite{Chen2022}. These high-cost samples are chosen with an active learning scheme~\cite{schran2020committee, krogh1994neural}, discussed in detail below. 
The main workflow for our approach is illustrated in Fig.~\ref{fig:fig1}. Our transfer learning framework is based on \citet{Chen2022}, which deploys transfer learning to optimize potential energy surfaces between different classical methods, including methods that are stochastic and thus require training on noisy data sets.  
%However, a main difference is that instead of transfer-learning energies from classical electronic structure calculations, we develop technique for transfer-learning of energies from VQE calculations performed on quantum computers. 
We integrate transfer learning to energies obtained with VQE with a query-by-committee and active learning approach \cite{schran2020committee,krogh1994neural}.

The difficulty of evaluating gradients of the energy on quantum hardware motivates our approach, which we demonstrate by employing transfer learning between DFT and VQE datasets for the water monomer and water dimer. We have chosen water monomers and dimers as illustrative examples because they are few-atom systems that manifest both intra- and intermolecular dynamics and because they can be used as starting points for developing high-accuracy water force fields of longstanding interest to the chemistry community \cite{distasio2014individual,gillan2016perspective,cheng2019ab,babin2013development,babin2014development,medders2014development,reddy2016accuracy,nandi2021ccsd,yu2022q} using quantum hardware. For our test systems, we show that we can make high-accuracy predictions of the potential energy surfaces using transfer learning by correcting DFT-trained neural networks with just tens of VQE points. We also demonstrate that these surfaces, trained on energies alone, can generate forces accurate enough to produce stable dynamics that reproduce radial distribution functions generated by more accurate benchmark methods.

\subsection{Training Data Generation}

To perform transfer learning, we first produce Density Functional Theory (DFT) and Unitary Coupled Cluster (UCC) data sets for training. These data sets contain different water monomer and dimer geometries and their accompanying energies.
\begin{figure*}
    \centering
    \includegraphics[width=0.9\linewidth]{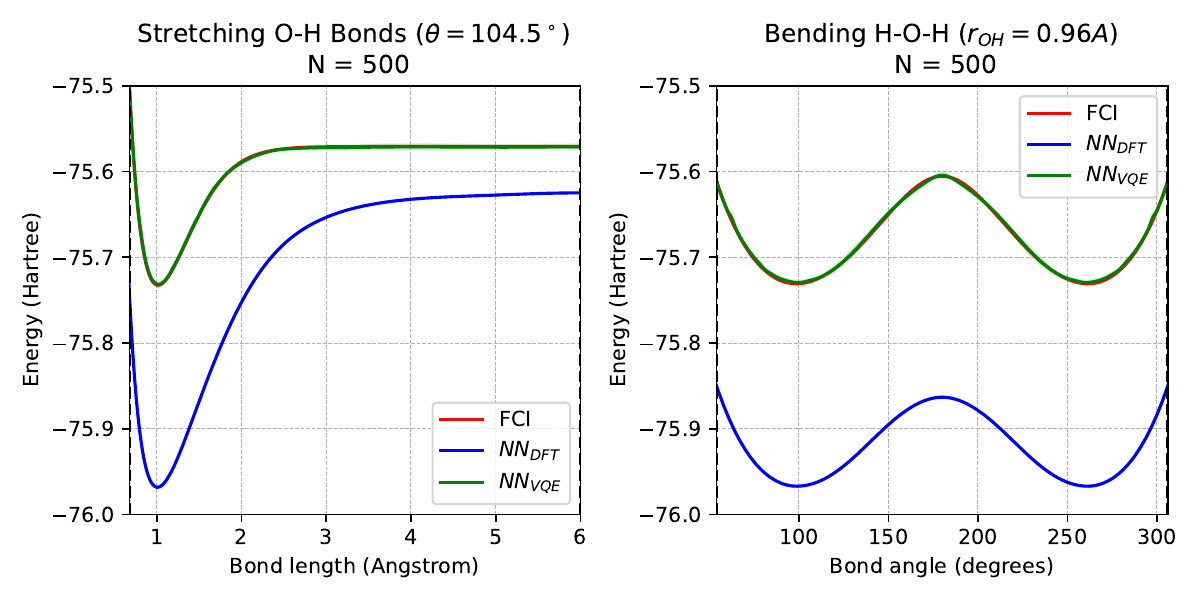}
    \caption{Potential energy surfaces of monomer H$_2$O. (Left) The potential energy as a function of one O-H bond length, fixing the bond angle at $\theta = 104.5^\circ$ and the other O-H bond length at 0.96 \AA. (Right) The potential energy as a function of the bond angle, fixing both O-H bond distances at 0.96 \AA. Both plots show the energies obtained from FCI (red), the DFT-only trained BPNN (blue), and the transfer-learned BPNN using VQE data (green). }
    \label{fig:pes_v4}
\end{figure*}
\subsubsection{DFT Data}

\paragraph{Water Monomer}
% Water monomer data distribution

We computed 18,623 water monomer configurations, including their molecular geometries and corresponding energies. Energies were calculated at the PBE0/STO-6G level of theory \cite{perdew1996generalized} using the Gaussian~\cite{g16} calculator in the Atomic Simulation Environment package~\cite{larsen2017atomic}. The STO-6G basis was chosen for its simplicity and computational efficiency: Water possesses 7 molecular orbitals in this basis, meaning modeling this molecule on quantum hardware would necessitate 14 qubits (one qubit for each spin-orbital). % which is within reach of modern hardware.  

The geometries were methodically sampled using a grid-based approach along the internal coordinates of water: water's two O-H bond lengths and its H-O-H angle. Given that the water molecule is one of the simplest molecular systems with a multidimensional potential energy surface (PES), grid-based sampling, which typically scales exponentially with the number of dimensions, is feasible for this system and ensures comprehensive coverage of water's potential energy surface. A detailed analysis of the distribution of O-H bond lengths and H-O-H bond angles in the computed dataset can be found in Appendix \ref{app:monomer dataset}. Our comprehensive sampling of the internal coordinates illustrates the robustness of our dataset, which ensures its reliability for modeling the global PES of water.

\paragraph{Water Dimer}

Our water dimer database consists of 64,061 dimer configurations. Each data point includes the geometric coordinates of the water dimer configuration and its corresponding energy. A water dimer has more atoms and electrons than a water monomer, leading to higher computational complexity and resource requirements. %Using a simpler basis set helps in reducing the time and resources needed for calculations.
The energies were calculated using the PBE0/STO-3G level of theory, which requires 14 molecular orbitals. 

To generate our water dimer database, we use a sampling method that begins with the global optimal configuration of the water dimer, wherein all degrees of freedom are constrained except for the oxygen-oxygen (O-O) distance, ensuring that the diversity of configurations primarily arises from variations in this parameter. A grid sampling approach is employed to explore configurations with O-O distances ranging from 2.0 to 8.0 angstroms, sampling a unique configuration at every 0.1-angstrom increment within this range. Each sampled configuration is then used as the initial frame for subsequent MD simulations carried out by Orca~\cite{Neese_2012,neese2020orca}.
The molecules' initial velocities are calibrated to reflect a temperature of 300 K, ensuring that the starting conditions represent room-temperature dynamics. Utilizing a timestep of 0.5 femtoseconds, these simulations capture the movements of atoms with high temporal resolution. Temperature regulation is achieved through the Canonical Sampling Velocity Rescaling (CSVR) thermostat set at 300 K with a time constant of 100.0 fs, providing a stable thermal environment over the course of 2000 steps per simulation. 

Finally, an additional molecular dynamics (MD) trajectory has been incorporated into the dataset, with its initial frame based on the global optimal structure of the water dimer. This 5 ps trajectory was simulated using the same parameters as the other MD simulations to ensure consistency across the dataset. Including this extra trajectory is strategic, aiming to enrich the dataset with ample data points around the equilibrium structure. This enhancement is crucial for providing a comprehensive representation of the water dimer's behavior in its most stable form, thereby facilitating the neural network's ability to learn and predict with greater accuracy. 
%This approach underscores our commitment to
Our methodology ensures that we are constructing a dataset that spans a broad range of configurations and emphasizes the critical regions of configuration space, such as the equilibrium structure, whose accurate representation is essential for producing meaningful molecular dynamics.

\subsubsection{UCCSD Data}

To create our higher-accuracy, ``quantum'' data set, we perform classical optimization of a parameterized quantum circuit 
%simulations of VQE 
using the Unitary Coupled Cluster (UCC) ansatz for the water monomer and dimer datasets. These calculations were performed using a sparse wavefunction circuit simulator~\cite{mullinax2023largescale}, which has been shown to efficiently obtain exact or approximate ground-state energies of quantum chemical systems. For the water monomer, we optimize the full UCC with singles and doubles (UCCSD) ansatz with no approximations within the STO-6G basis set. 
%This amounts to a VQE simulation using 14 qubits.
For the water dimer, we optimize the UCCSD ansatz but only keep 400 single and double-excitation operators. During the optimization procedure, only 10,000 amplitudes are kept for the sparse simulator~\cite{mullinax2023largescale}. The truncation of the ansatz and the number of amplitudes of the wavefunction define this approximate VQE procedure. Still, it yields energy values that are virtually exact for this system and thus adequate for demonstrating the transfer learning procedure. We illustrate the accuracy in Appendix~\ref{app:vqe_error}, presenting a histogram of the absolute energy errors between the UCCSD circuit energies and those obtained from FCI. We optimize the UCCSD ansatz using the STO-3G basis set, which amounts to a 24-qubit simulation after we freeze two core orbitals.
\begin{figure}
    \centering
    \includegraphics[width=\linewidth]{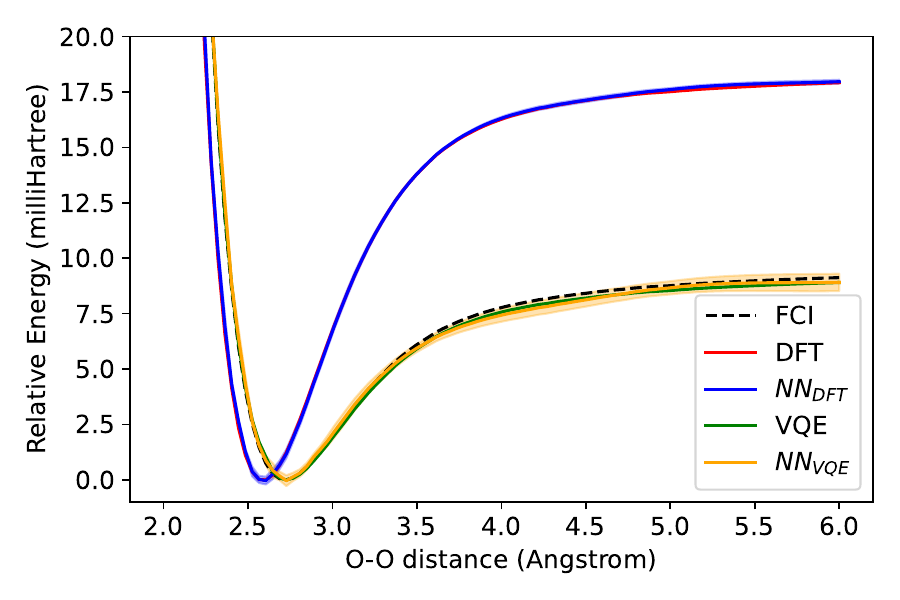}
    \caption{Potential energy surface of the water dimer
    as a function of the intermolecular O-O distance holding the geometries of the individual water molecules fixed. The energies at the global minima of the curves are subtracted away to facilitate comparison. The minima are located at $E_{DFT} = -150.5139$ Ha, $E_{FCI} = -150.0376$ Ha, and $E_{VQE} = -150.0370$ Ha. }
    \label{fig:PES_dimer}
\end{figure}

\subsection{Model Training}\label{sec:modtrain}
To perform transfer learning, we train a Behler-Parrinello neural network (BPNN) architecture~\cite{PhysRevLett.98.146401}, which provides the energies and forces of the molecules as a function of their atomic positions. We use the set of symmetry functions previously used for water in Ref.~\cite{morawietz2016how} within n2p2 \cite{singraber2019library,singraber2019parallel}. While this network allows for training with both energies and forces, our training data consists only of the molecular geometries (configurations) and the DFT or UCCSD circuit energies. Here, we employ a committee machine learning procedure \cite{schran2020committee} in which we train multiple models simultaneously; each model is trained on a random split of the larger dataset into training and validation sets. Then, given a configuration, the resulting energy is a weighted average over all model evaluations. Given a configuration, the committee's output gives us the mean energy among all models and the variance. The variance provides a quantitative measure of how uncertain that model is about the energy of the given configuration, which is beneficial to understanding how uncertain the model is when evaluating geometric configurations on which the model has not been trained. %This measurement is crucial in the active learning procedure we conduct.

To start, we train 8 BPNNs with the large datasets obtained via DFT calculations. These neural networks are initialized with random weights and trained on 90\%-10\% training-validation splits for 25 epochs, each BPNN having a different 90-10 split. This ensures that no two BPNNs are identical. Once training with the DFT dataset is complete, we continue training the eight models, but now only with energy data obtained from the UCCSD circuit. We employ an active learning procedure for both the water monomer and dimer datasets to iteratively add training points to the overall dataset. Because the cost of obtaining VQE energies is high, the active learning procedure ensures that every training point is chosen to facilitate an efficient training process. 

For both the water monomer and dimer, the transfer learning procedure is as follows. We start by sampling 20 random configurations from the DFT dataset. We train the committee after obtaining the corresponding UCCSD circuit energies with those configurations. We again use a 90-10 train-validation split and train for 25 epochs. Once training is complete, we choose the epoch weights that yield the lowest validation loss. This is crucial, as with transfer learning, the minimal number of training points makes it easier for the model to over-fit. With the new weights for the models, we then predict the energies of all the configurations in the DFT dataset. For each configuration in the dataset, the committee of BPNNs will give a mean energy \emph{and} a variance, which measures the committee's uncertainty about the energy. We add the 20 configurations with the highest variance into the training set and perform new 90-10 train-validation splits for the next training iteration, where we restart training from the DFT-trained weights. The process of training and adding data into the training set is repeated until we reach a desired convergence. There are two stopping criteria that we can employ. First, we can stop when the variance in all the evaluated energies is below some threshold, implying that the model has reached a level of certainty over all the data points it has seen. Secondly, if we are given a test set the model has not trained on, we can stop when the test loss is below some threshold.

For the water dimer, the DFT dataset is obtained via grid sampling and DFT-driven molecular dynamics. In addition to sampling from this DFT dataset in the same fashion as we sampled for the monomer, we also perform and then sample from molecular dynamic simulations driven by the trained BPNNs. We do this because, in contrast with the water monomer dataset, the configuration space is too vast to explore. Therefore, we also concentrate on configurations we expect to see in the dynamics. Sec.~\ref{sec:md} describes how we perform MD simulations. Our MD-driven active learning procedure is as follows. We start by performing the transfer learning procedure described above by sampling from just the DFT dataset until the uncertainty evaluated over the whole dataset is below chemical accuracy. Then, we stop sampling from the DFT dataset and instead sample from MD simulations driven by BPNN. These MD simulations provide us with a trajectory of geometric configurations, which we input into the transfer-learned BPNNs. Like the above, we add the 20 points in the trajectory that yield the highest variance. To prevent cross-correlation between sampled points, we ensure that points chosen from the trajectory are no less than 100 fs apart. We iterate this procedure again until we reach a converged error.

\subsection{Molecular Dynamics}\label{sec:md}

We performed molecular dynamics simulations on our water monomer and dimer on a UCCSD(VQE)-quality potential energy surface and surfaces produced by other levels of theory for comparison. While these species have too few degrees of freedom to be properly thermostatted and thus yield meaningful equilibrium dynamics, we can nonetheless numerically perform MD simulations that can give us insights into the accuracy and stability of our transfer-learned neural networks. To do so, we use LAMMPS as our MD engine~\cite{thompson2022lammps,singraber2019library}. We start from a near-equilibrium geometric configuration and run the MD engine with a 0.5 fs time step at a temperature of 300 K using the Nose-Hoover thermostat within an NVT ensemble with $\tau= $ 1 ps. We run this for 5 ps to let the molecule(s) equilibrate. With the last frame of that run as the new initial configuration, we start a new MD simulation with an NVE ensemble with the same time step but for 50 ps, and after equilibration, compute observables such as atom-atom distances.

\section{Results \label{sec:res}}

\subsection{Training with DFT}
As a first step toward testing our BPNNs, we begin by testing the accuracy with which we can learn our DFT data. In Table~\ref{tab:dft}, we show the training, validation, and test loss for our DFT-trained BPNNs for the water monomer and dimer. These are calculated by averaging the mean absolute error between the BPNN and DFT-reported energies. With both the monomer and dimer, the DFT-trained BPNNs produce nearly identical training and validation loss values with sub-chemical accuracy, highlighting the accuracy of our fits and the absence of overfitting. Additionally, we evaluate both models on test sets that are completely separate from the training data. For the monomer, we sampled 5,000 points from the configuration space. For the dimer, we sampled 5,000 points from an MD simulation using the same procedure as Sec.~\ref{sec:md}. While larger than the training and validation losses, the test loss still yields similar loss values, indicating strong generalizability of the models.
\begin{table}[]
    \centering
    \begin{tabular}{c|c|c}
       & monomer & dimer\\
\hline
train      & 1.137  & 0.089 \\ 
validation & 1.147 &  0.091\\
test       & 1.706 & 0.157
\end{tabular}
    \caption{Mean absolute errors (MAE) in mHa of the DFT-trained BPNNs for the water monomer and dimer over different datasets.}
    \label{tab:dft}
\end{table}

% \begin{itemize}
% \item Add figure comparing DFT and transfer-learned DFT potentials
% \item Discuss how many epochs/transfer learning rounds you need for convergence
% \item 
% \end{itemize}

\subsection{Transfer Learning with VQE}
\begin{figure}
    \centering
    \includegraphics[width=\columnwidth]{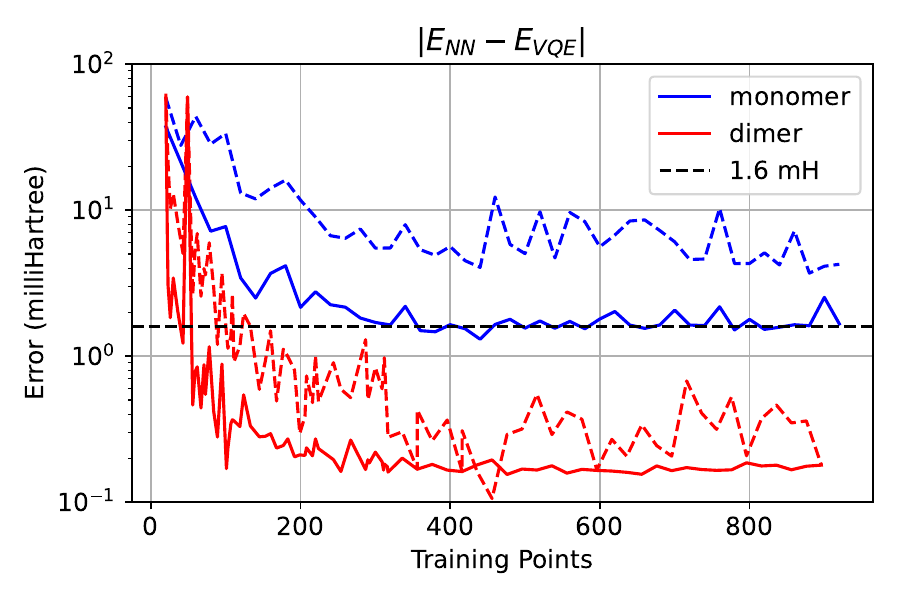}
    \caption{Mean average error (MAE) as a function of the number of VQE training points $N_\text{TL}$ used in transfer learning for the water monomer (blue) and dimer (red). The dashed lines represent the loss of the test set at each step, which consists of 20 points. These points are added to the training set during the next iteration. The solid lines represent the MAE over all the previously sampled training points. The black dashed line depicts chemical accuracy.}
    \label{fig:update_error}
\end{figure}

With the BPNNs well-trained on the DFT data, we employed transfer learning to train on the UCCSD(VQE) data. In this process, we also use an active learning procedure (see Sec.~\ref{sec:modtrain}) to iteratively add points to the training dataset to be as data-efficient as possible, given the more significant expense associated with acquiring the UCCSD(VQE) data. Data efficiency also constrains us in developing a test set to evaluate our models, which is essential for determining how successfully the model has been trained. We overcome these constraints by strategically using UCCSD(VQE) data for testing and training. At each iteration in the active learning procedure, we obtain at most 20 configurations that yield the highest committee disagreement among the BPNNs. These configurations at each iteration serve as a practical test set. After calculating the UCCSD energies on these configurations, but \emph{before} adding these points into the training dataset, we evaluate the trained model with these points to see how inaccurately the network predicted their UCCSD(VQE) energies. Furthermore, this test loss error is biased to be high because the 20 points used in this test set are the configurations where the model performs the worst, i.e.,  where the models disagree the most and, hence, likely where the model is most inaccurate. Figure~\ref{fig:update_error} shows this loss as a function of the number of training points used. Unsurprisingly, the test error decreases as more data points are added to the training set. The tradeoff shown in Fig.~\ref{fig:update_error} provides a way to choose what level of accuracy we want our model to have by selecting the number of points needed to reach the level of accuracy desired.

In addition to the test loss, Fig.~\ref{fig:update_error} also shows the Mean Average Error (MAE) over all of the VQE data on which the model was trained. This loss and the test loss provide general upper and lower bounds on the error over the space of geometric configurations of the respective molecules. 

For the water monomer, the test loss flattens out at around 5 milliHartree, while for the dimer, the test loss flattens at around 0.3 milliHartree. These values are reasonable given the MAE from the DFT training (see Table~\ref{tab:dft}). The monomer has a particularly elevated MAE because of highly distorted and thus higher energy monomer geometries in the data set.

\subsection{Potential Energy Surfaces}

Based on Fig.~\ref{fig:update_error}, the energy error does not improve if we train with more than 400 points. We can, therefore, view the energy as converged and look at potential energy slices along different reaction coordinates to evaluate the trained models. Figure~\ref{fig:pes_v4} illustrates the potential energy surface of monomer H$_2$O using 500 VQE training points. We show the PES along two reaction coordinates: one in which the O-H bond length is varied and one in which the bond angle is varied. The PESs constructed from the transfer-learned models lie precisely on top of the FCI-constructed PES, which is the exact solution within the given basis set.

We also find similar results when we plot the PES for the water dimer, for which we fix the internal geometries of the waters to their global minima but vary the O-O distances between the water molecules, translating them diametrically towards or away from each other. Figure~\ref{fig:PES_dimer} shows this transfer-learned PES constructed using 400 UCCSD(VQE) training points and compares it with the DFT, exact VQE, and FCI PESs. From this plot, one can observe that the DFT-level PES is different from the UCCSD/FCI-level PES, and with just 400 VQE training points, the model was able to translate from the lower DFT theory to the higher-level UCCSD/FCI theory.

Combined, our water monomer and dimer results illustrate that our transfer learning procedure can reproduce both intramolecular motions, as exhibited in the monomer test case, and often lower energy intermolecular motions, as shown in the dimer test case. For the water dimer, we did not focus our training as much on intramolecular forces as we did for the water monomer case, where we added configurations of highly distorted geometries. In general, for this work, we are only interested in configurations that would be realistically seen in a molecular dynamics simulation. Given the sizeable geometrical landscape of the water dimer, it would be far too costly to include the many but highly improbable distorted intramolecular geometries. In contrast, the water monomer's geometric landscape is small enough to include highly distorted geometries that one would not expect to see in MD simulations.

\begin{figure}
    \centering
    \includegraphics[width=\linewidth]{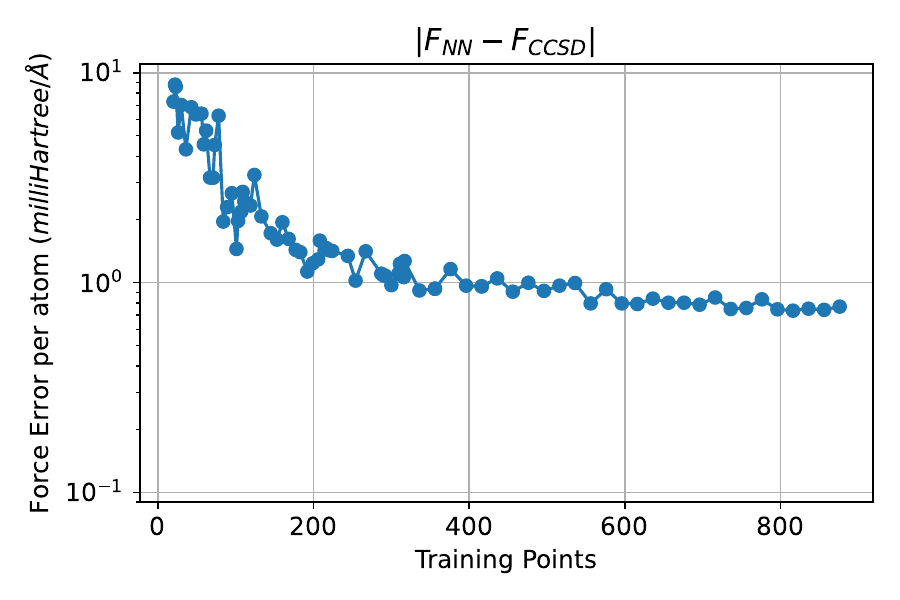}
    \caption{Mean force error between the UCCSD-BPNN and CCSD forces per atom for the water dimer. We sample 500 configurations from an MD simulation and compute the mean norm difference between the forces obtained from the UCCSD-BPNN and CCSD calculations.}
    \label{fig:force_error}
\end{figure}

Finally, to evaluate the performance of our models on accurately computing forces from the NN-PES for MD simulations, we compare them to forces obtained from classical CCSD. For the water dimer, we sample 500 configurations from an MD trajectory and compute the L2 norm of the difference between the forces obtained from the UCCSD-BPNN and forces from classical CCSD calculations~\cite{zhang2022differentiable, sun2020recent}. We show how this error decreases as the BPNN gains more VQE training points in Fig.~\ref{fig:force_error}. We find that the force per atom converges to $\sim$0.77 milliHartree/\AA ~(0.021 eV/\AA). This error is comparable to those observed in Ref.~\cite{Cancan2022}.

\subsection{Accuracy and Stability of the Molecular Dynamics}
With the BPNNs trained on DFT and DFT+UCCSD data, we test the accuracy and stability of molecular dynamics simulations run on the potential energy surfaces from these models. We specifically look at the dynamics of the water dimer. We ran 50 ps simulations of the water dimer at 300 K with a 0.5 fs time step. This MD simulation was performed four times, with each iteration obtaining energy gradients from either the true DFT, DFT-BPNN, CCSD, or UCCSD-BPNN potentials. Firstly, at a qualitative level, simulations run using all of these potentials were stable, with the monomers remaining bound and interacting with one another over the entire duration of the trajectories. A more quantitative way to measure the stability and accuracy of these BPNN models is by comparing the distributions of the inter- and intra-atomic bond distances throughout their MD simulations.

\begin{figure}
    \centering
    \includegraphics[width=\linewidth]{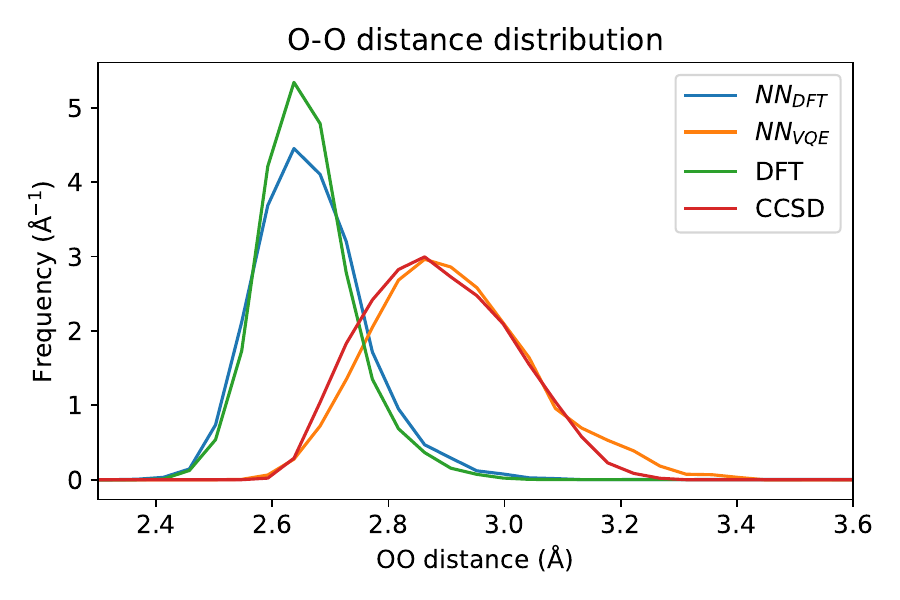}
    \includegraphics[width=\linewidth]{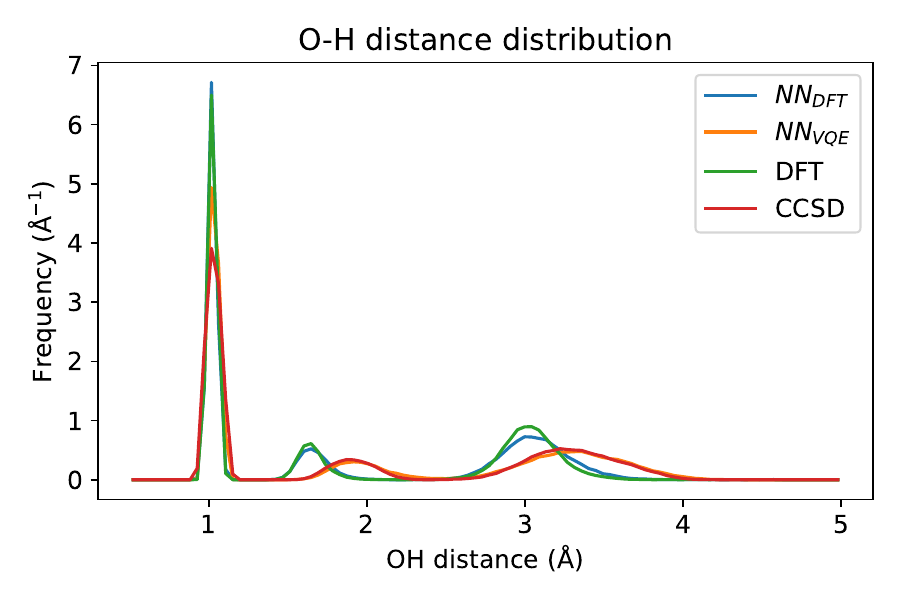}
    \caption{(Top) O-O distribution and (Bottom) O-H distribution of the water dimer over a 50 ps MD simulation performed with a 0.5 fs timestep.}
    \label{fig:dimer_dist}
\end{figure}

% distributions and overlaps
Figure~\ref{fig:dimer_dist} shows histograms of the O-H and O-O bond lengths over the 50 ps MD trajectories. We find great agreement between the DFT-driven distributions and the NN-DFT distributions. We see even better alignment between the CCSD-driven distributions and the NN-UCCSD-driven distribution. These distributions are a sensitive measure of the quality of our potentials and suggest that their energies and forces - which we did not explicitly train on - are both sufficiently accurate to drive meaningful MD simulations.

\section{Conclusions \label{sec:conc}}

%In this manuscript, we have described a new approach for performing molecular dynamics simulations using quantum hardware. 
This work demonstrates an approach for performing molecular dynamics using quantum processors that we anticipate will facilitate the high-accuracy modeling of the molecular dynamics of complex biomolecules and molecular catalysts on quantum hardware in the near future. 
Since it is challenging to obtain forces on quantum processors directly, we train neural networks using energies from quantum hardware and leverage those neural networks to predict both molecular energies and forces. We employ a combination of transfer and active learning to make the most efficient use of the limited number of energies we obtain from quantum algorithms. We first learn from cheaper and more abundant DFT energies and then correct them using Unitary Coupled Cluster data. With this approach, we can model the internal dynamics of the water monomer and both the inter- and intramolecular dynamics of the water dimer. The accuracy of our potential energy surfaces and forces are corroborated by our ability to reproduce distance distribution functions of O-O and O-H distances with minimal errors. Although we did not explicitly illustrate these capabilities in this manuscript, our technique can also be readily employed to perform other tasks that involve forces, including geometry relaxation. 

Even though this approach has proven successful on systems of small molecules, it could be improved in various ways to accommodate larger molecules of greater scientific interest. First and foremost, we have employed Behler-Parrinello Neural Networks (BPNNs). While BPNNs remain powerful networks, recent research suggests that equivariant message-passing neural networks such as MACE \cite{batatia2022design,batatia2022mace} and Allegro \cite{musaelian2023learning} may offer clear performance advantages by replacing BPNN symmetry functions with machine-learned equivalents \cite{kyvala2023optimizing}. Though our preliminary studies show BPNNs are on par if not better than MACE for the water datasets we have considered (see Appendix Sec.~\ref{app:mace_results} for a thorough comparison), these networks may perform better on larger systems and at predicting reactive dynamics involving bond breakage and formation with which fixed descriptor sets will have difficulty, enabling the modeling of synthetic and biomolecular catalysis \cite{hu2024training}. We would additionally need to improve the efficiency with which we sample training points to train such networks, which are often defined by many more parameters. In this work, we relied upon molecular dynamics simulations and, in some cases, sampling normal modes to obtain training points that traverse the relevant potential energy landscape, leading to comparatively large training sets of thousands of points. While such large data sets may help train more sophisticated architectures with more parameters, many configurations are likely correlated. Thus, the errors we obtain could likely be improved by data sets either specifically tailored to the chemistry in mind or with more active sampling of the initial training set. 

One reality of computing on modern NISQ hardware that we have overlooked by simulating on classical hardware is the noise that would inevitably accompany the UCC energies obtained on quantum hardware. While such errors may pose a challenge for interpreting single-point calculations (e.g., as used in simulations of catalysts), previous efforts in learning force fields from stochastic electronic structure data (e.g., from quantum Monte Carlo algorithms) have shown that most machine learning algorithms are largely immune to this noise \cite{Cancan2022,niu2023stable}. Assuming enough data and reasonable error bars, machine learning algorithms can readily learn stable force fields that have smaller single-point energy uncertainties than the original data \cite{ceperley2024training}. Our approach of combining quantum data with classical machine learning algorithms can thus be viewed as a fruitful marriage that makes the most of both techniques.

Natural extensions of this work involve scaling to larger, more reactive systems, including reactions involving either homogeneous or inhomogeneous catalysts. As alluded to above, this will necessitate developing more efficient data generation, sampling, and training procedures. The modeling of reactive systems will also necessitate dataset generation and model architectures that can readily model bond breakage and formation. In cases in which proton transfer is involved, our techniques would also have to be generalized to accommodate nuclear quantum effects. We view these challenges as motivating but not limiting, given the pace of current algorithmic developments, and thus believe that this work lays out a novel yet lasting blueprint for how molecular dynamics can be performed using quantum hardware. This vital task has many potential impacts on modeling condensed phases and biological systems. 

\section{Data and Code Availability}

The DFT and VQE datasets for the water monomer and dimer used in this work may be found at \href{https://doi.org/10.5281/zenodo.11176825}{10.5281/zenodo.11176825} Our transfer learning codes may be available upon request after going through the NASA release process.

\section{Acknowledgements} 
 We thank Gopal Iyer for insightful discussions. NMT, AK, GMR, CHB, PV, YP, and BR's research was supported by Wellcome Leap as part of the Quantum for Bio Program. AK acknowledges support from USRA NASA Academic Mission Services under contract No. NNA16BD14C through participation in the Feynman Quantum Academy internship program. BKC acknowledges support from the NSF Quantum Leap Challenge Institute for Hybrid Quantum Architectures and Networks (NSF Award 2016136). MSC acknowledges support from the Simons Foundation (Grant No. 839534). Electronic structure calculations were conducted on classical hardware using computational resources and services at the Center for Computation and Visualization, Brown University.
 This research used resources from the National Energy Research
Scientific Computing Center, a DOE Office of Science User Facility supported by the Office of Science of the US Department of Energy under Contract No. DE-AC02-05CH11231 using NERSC award ASCR-ERCAP0024469.

\appendix

\section{Water Monomer Dataset}
\label{app:monomer dataset}

\begin{figure}
    \centering
    \includegraphics[width=\columnwidth]{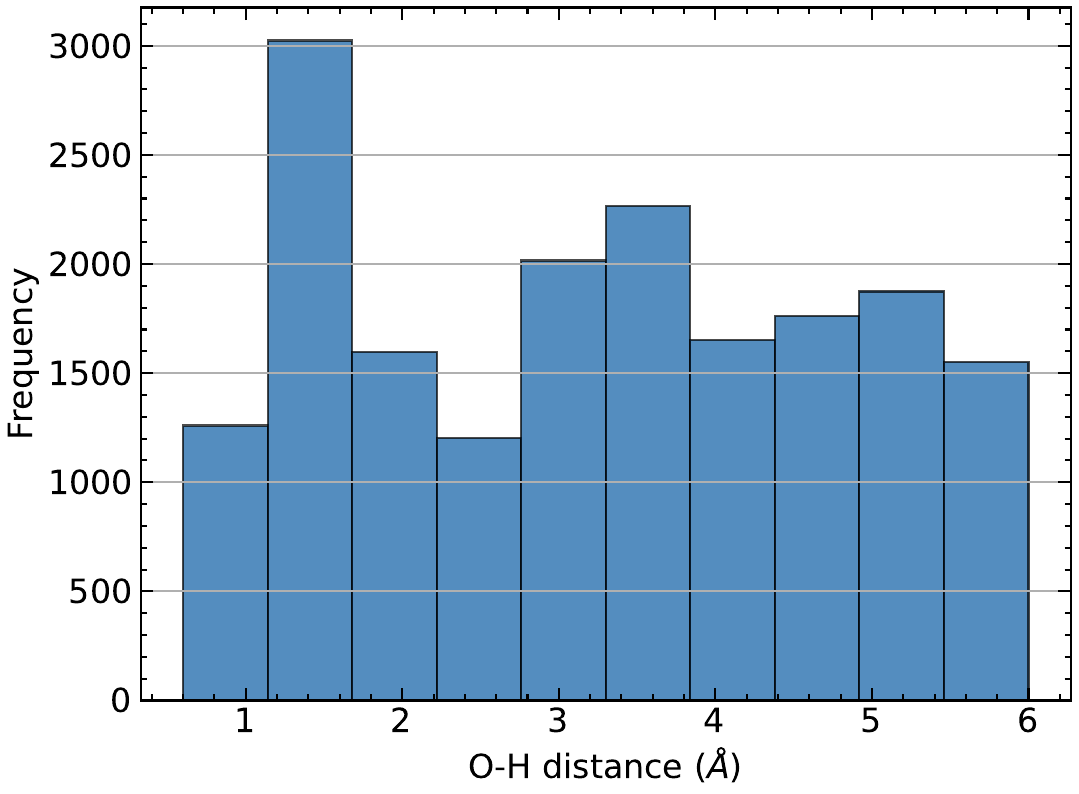}
    \includegraphics[width=\columnwidth]{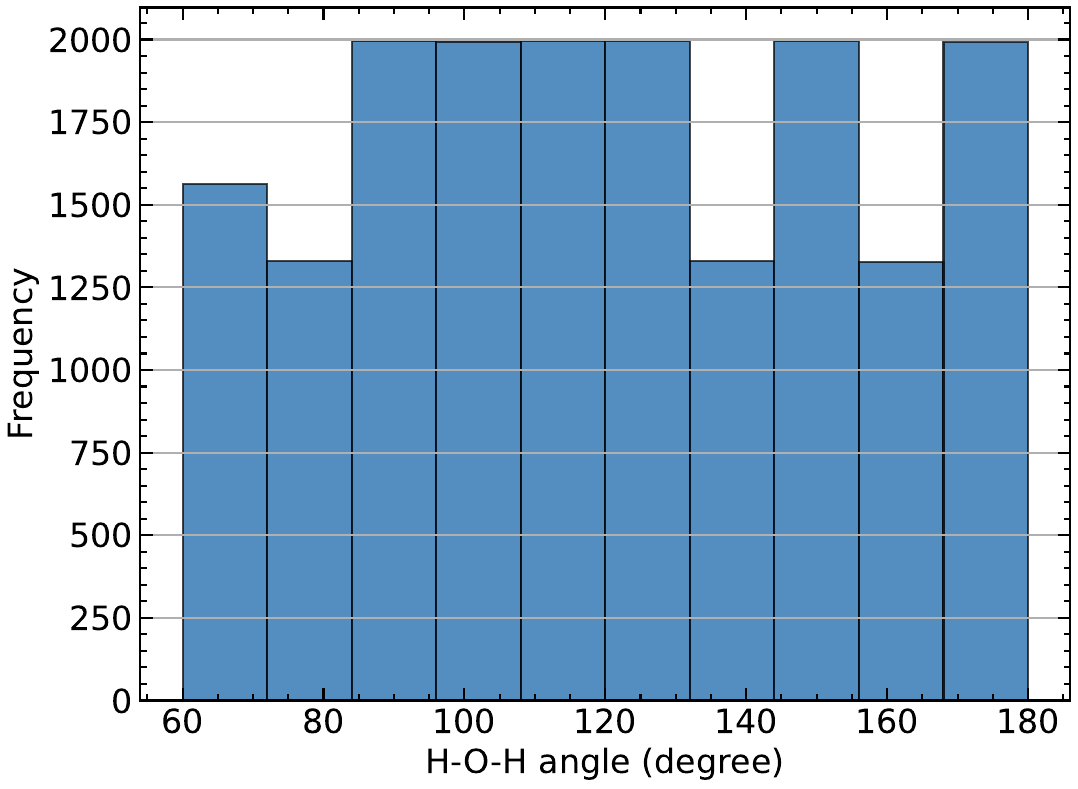}
    \caption{Histograms of the frequencies of different water angles and bond lengths in the water monomer dataset. (Top) The frequency distribution of O-H bond lengths in the computed dataset which peaks around 1.016 \AA, the equilibrium bond length at the PBE0/STO-6G level of theory. (Bottom) Frequency distribution of H-O-H bond angles in the computed dataset, highlighting the prevalence of configurations around 97.27$^{\circ}$, water's equilibrium bond angle at the PBE0/STO-6G level of theory.}
    \label{fig:monomer_hist}
\end{figure}

The histogram in Fig.~\ref{fig:monomer_hist}(top) presents the distribution of the O-H bond lengths present in our monomer dataset. This distribution ranges from 0.9 to 6 \r{A} and peaks between 0.9 and 1.5 \r{A}, around the equilibrium bond distance estimated to be 1.016\r{A} at the PBE0/STO-6G level of theory. This spread indicates an extensive sampling of molecular geometries with equilibrium to significantly elongated bond lengths that capture both low-energy stable configurations and higher-energy dissociative states.

The distribution of the H-O-H bond angles is displayed in Fig.~\ref{fig:monomer_hist}(bottom). The angle frequencies peak around $97.27^{\circ}$, corresponding to the water molecule's equilibrium configuration at the PBE0/STO-6G level of theory. The distribution covers a wide range of angles from approximately $60^\circ$ to $180^{\circ}$, allowing for the investigation of both linear and bent molecular geometries.

\section{Water Dimer Dataset}
\label{app:dimer dataset}

\begin{figure}[ht!]
    \centering
    \includegraphics[width=.8\columnwidth]{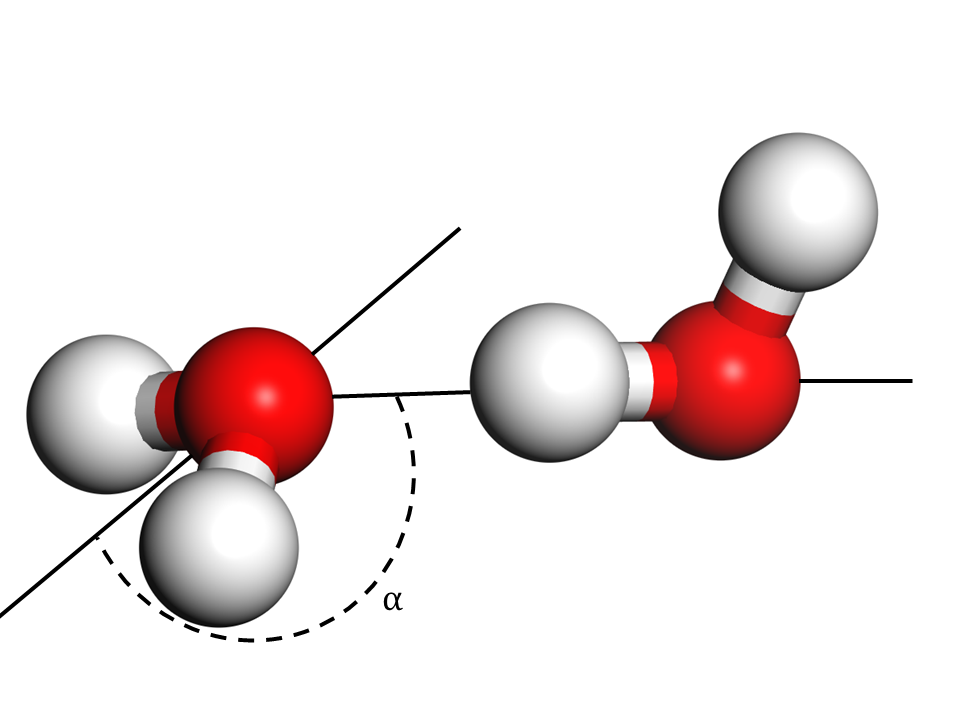}
    \caption{The equilibrium configuration of the water dimer at the PBE0/STO-3G level of theory. Red sphere: Oxygen atom; White sphere: Hydrogen atom; $\alpha$: The angle between the molecular plane of a water molecule and the line defined by the oxygen atoms of the dimer.}
    \label{fig:waterdimer}
\end{figure}

\begin{figure}[ht!]
    \centering
    \includegraphics[width=\columnwidth]{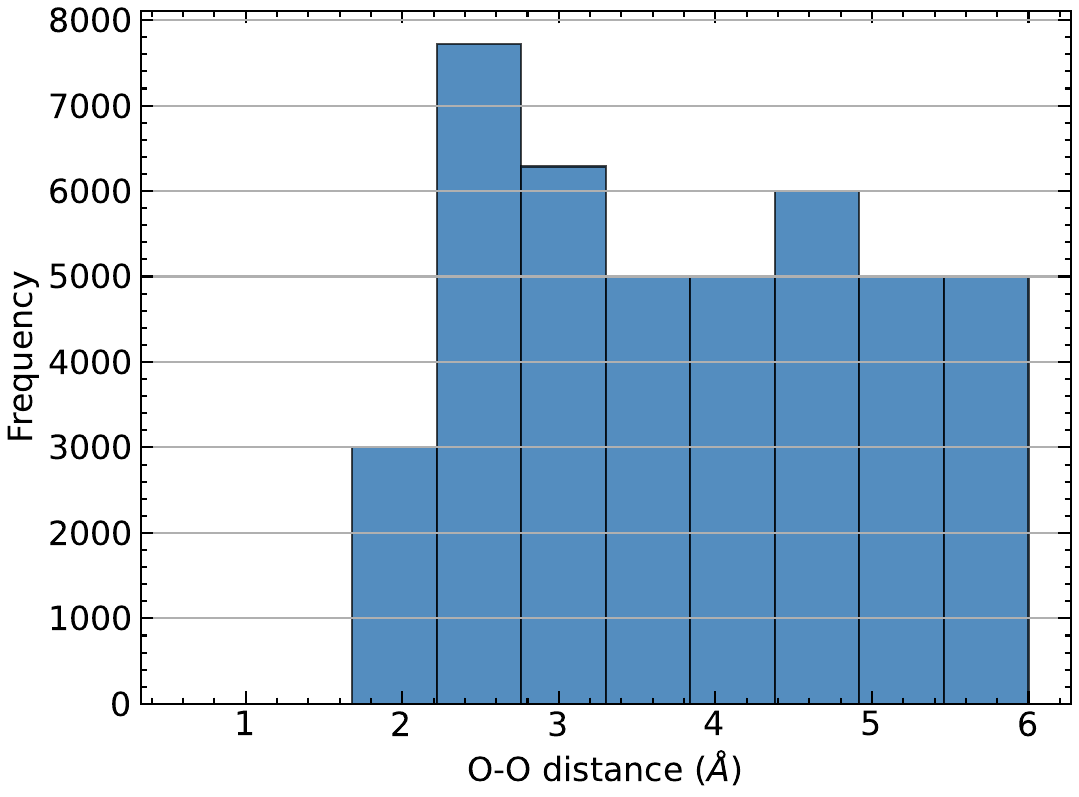}
    \includegraphics[width=\columnwidth]{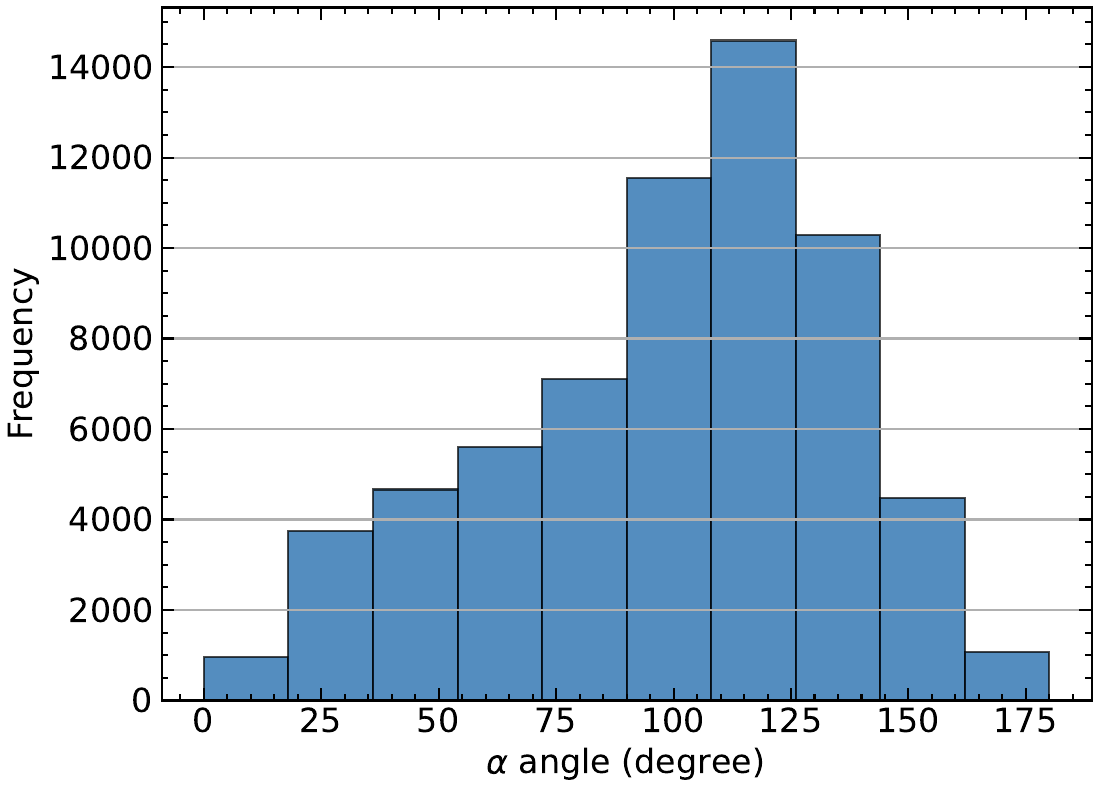}
    \caption{Histograms of the frequencies O-O bond distances and angles $\alpha$ in the water dimer dataset. (Top) Frequency distribution of O-O bond distances in the dataset, which peaks around the equilibrium distance of 2.64 \AA\  at the PBE0/STO-3G level of theory. (Bottom) Frequency distribution of $\alpha$ angles in the computed dataset. In the equilibrium structure at PBE0/STO-
    3G level of theory, the $\alpha$ angle is approximately 101$^{\circ}$.}
    \label{fig:waterdimer_histograms}
\end{figure}

Figure~\ref{fig:waterdimer} depicts the equilibrium structure of the water dimer and the related bond lengths and angles compiled in our dataset. 
Figure~\ref{fig:waterdimer_histograms} shows histograms for the O-O frequency distribution (top) and $ alpha $ (bottom), which is the angle between the plane of a water molecule and the radial vector between the two oxygen atoms.

\section{Water Dimer VQE Errors}\label{app:vqe_error}
\begin{figure}
    \centering
    \includegraphics[width=\columnwidth]{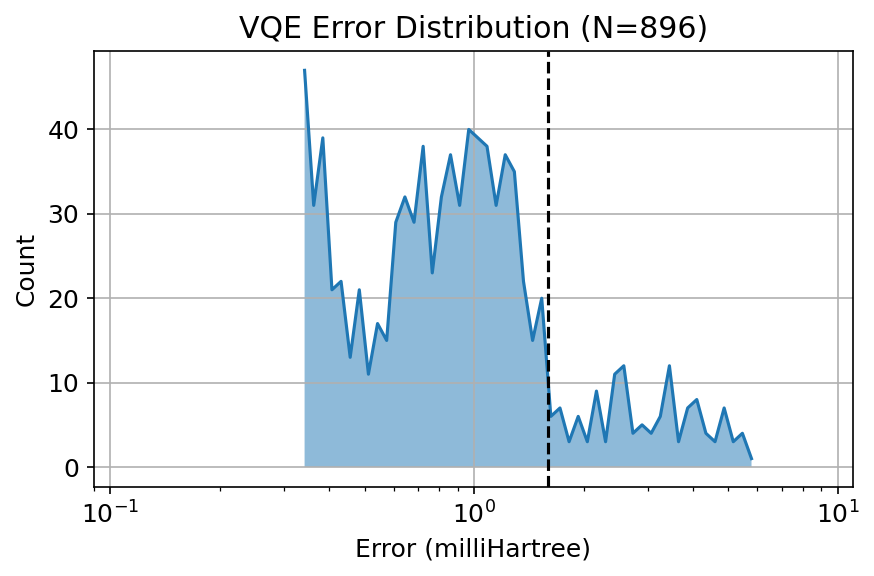}
    \caption{Histogram of the energy errors of the VQE energies relative to the FCI energies.}
    \label{fig:vqe_error}
\end{figure}

Figure~\ref{fig:vqe_error} presents the histogram of the (log) energy error of the VQE energies relative to the FCI energies of the water dimer. The VQE was performed with approximations as described in Sec.~\ref{sec:methods}. We find that most of the VQE energies used in the transfer learning are below chemical accuracy, with sub-milliHartree precision.

\section{MACE Results}\label{app:mace_results}

% Going to be using this as a temporary staging ground for MACE results

\begin{table}[h!]
    \centering
    \begin{tabular}{c|c|c|c}
       & monomer & dimer & dimer w/ forces\\
\hline
train      & 0.647 & 0.316 & 0.0129 \\ 
validation & 0.623 &  0.317 & 0.0129 \\
test       & 1.484 & 1.332 & 0.0229 
\end{tabular}
    \caption{Mean absolute errors (MAE) in milliHartree of the DFT-trained MACE NNs for the water monomer and dimer over different datasets. Averages and errors are over 8 independent neural networks.}
    \label{tab:mace}
\end{table}

In this section, we perform studies using MACE ~\cite{batatia2022mace,batatia2022design} on the monomer and dimer datasets in order to analyze how the quality of our predictions may differ if we replace our BPNN neural network architecture with more modern networks. MACE is an equivariant message-passing neural network, which, in comparison to BPNNs, learns spherical tensor features that efficiently incorporate many-body behavior through tensor products of the features.
The MACE energy is obtained by summing an energy readout function on each atom that uses the rotationally invariant features generated in every layer of MACE.

For MACE, we use a radial cutoff of 12 \AA, $ 128 $ channels, and maximal message equivariance of $ 2 $.
The MACE code is modified to allow for energy-only training available online~\cite{batton2024mace}.
Eight MACE models are trained with different initialized random weights and 90-10 training-validation splits for 7000 epochs for the monomer dataset and 1000 epochs for the dimer dataset.
Additionally, separate MACE models are trained on the dimer dataset with both energies and forces obtained from DFT for 500 epochs.
The mean absolute error in the energy over the training, validation, and test sets are shown in Table~\ref{tab:mace}.
The MACE models perform slightly better than BPNNs for the monomer dataset.
On the dimer dataset, the models trained on energies alone performed worse than BPNNs, while the models trained on both energies and forces performed better than the BPNNs. Note that the BPNNS are trained only on the energies.
The increase in performance by training on the forces is likely due to the forces providing local descriptors that help form the descriptors MACE learns during training, in comparison to BPNNs, which use fixed descriptors that provide an inductive bias.

\begin{figure}
    \centering
    \includegraphics[width=\columnwidth]{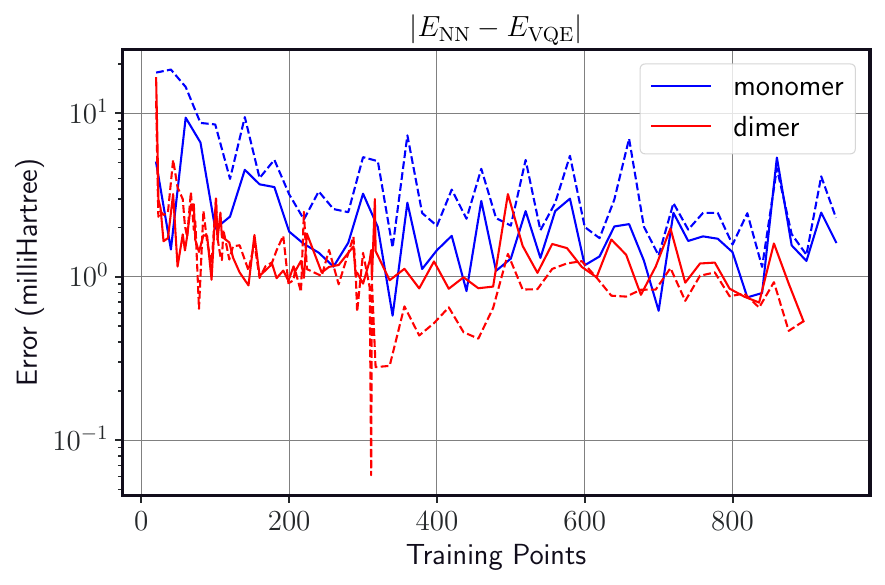}
    \caption{Mean average error (MAE) as a function of the number of VQE training points $N_\text{TL}$ used in transfer learning for the water monomer (blue) and dimer (red) using MACE with the same dataset used in Fig.~\ref{fig:update_error}. The dashed lines represent the loss of the test set at each step, which consists of 20 points. These points are added to the training set during the next iteration. The solid lines represent the MAE over all of the previously sampled training points.}
    \label{fig:update_error_mace}
\end{figure}

\begin{figure}
    \centering
    \includegraphics[width=\columnwidth]{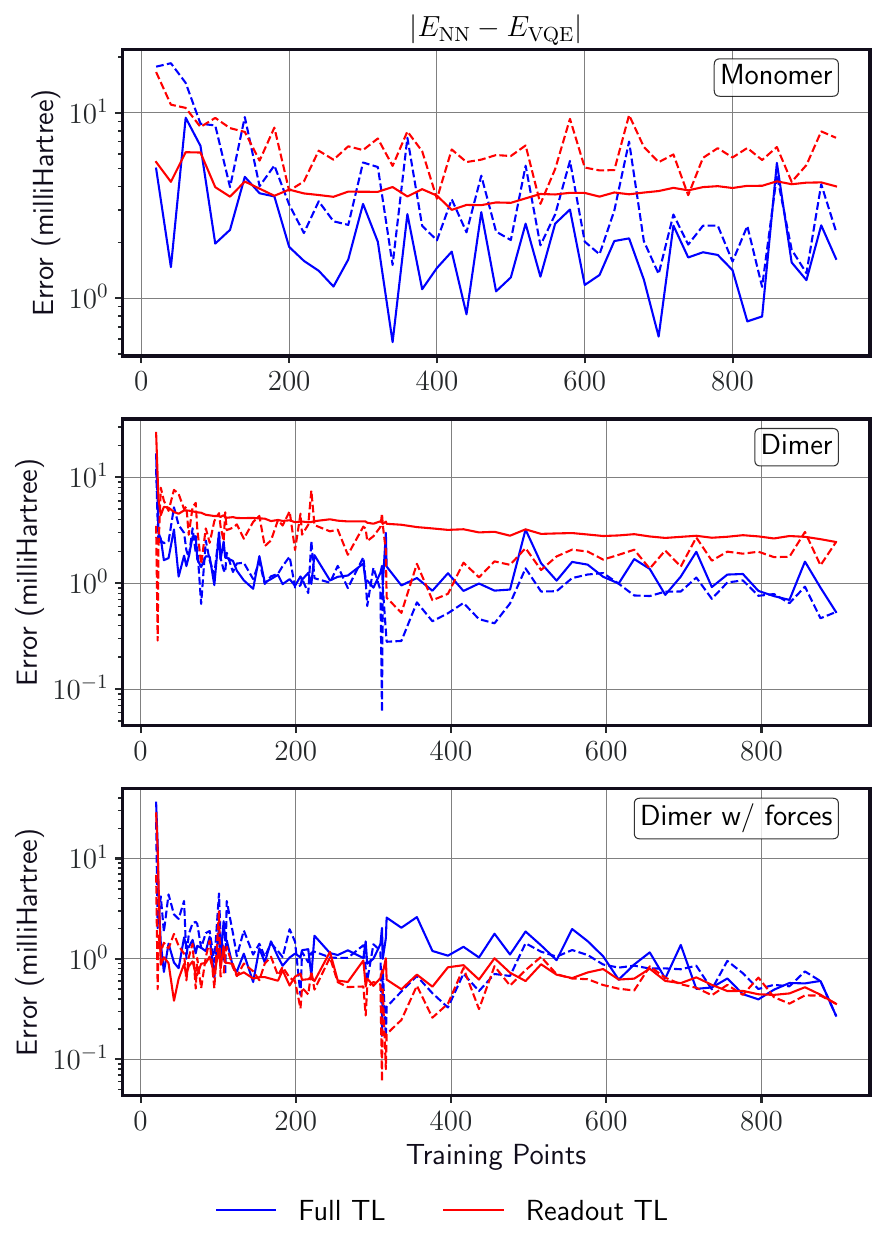}
    \caption{Mean average error (MAE) as a function of the number of VQE training points used in transfer learning, $N_\text{TL}$, for the water monomer and dimer using MACE where the whole model (blue) or only the energy readout (red) is retrained with the same dataset used in Fig.~\ref{fig:update_error}. The dashed lines represent the loss of the test set at each step, which consists of 20 points. These points are added to the training set during the next iteration. The solid lines represent the MAE over all of the previously sampled training points.}
    \label{fig:update_error_mace_ft}
\end{figure}

Transfer learning to the VQE data is done for the MACE models through the datasets used for the BPNNs per Fig.~\ref{fig:update_error}.
The corresponding loss and test curves are shown in Fig.~\ref{fig:update_error_mace} for the MACE models initially trained on energies alone for both the monomer and dimer, where we train with 5000 epochs for each iteration.
An additional study was done by optimizing the energy readout alone, shown in Fig.~\ref{fig:update_error_mace_ft}, along with results from optimizing the full architecture.
In the case of the monomer dataset, MACE performs on par with BPNNs. For the dimer dataset, MACE performs worse than BPNNs, with the MACE model trained on DFT energies and forces performing better than on DFT energies alone.
Optimizing just the energy readout reduces the accuracy, except for the MACE models trained on energies and forces for the dimer dataset.
This is indicative of the descriptors of the MACE model trained on both energies and forces being of higher quality than those trained on energies alone.

The MACE models' test error for the monomer dataset is greater than the training error for most iterations, while for the dimer dataset, the test error is less than the training error for most iterations.
Hence, the points that are maximally informative to the BPNNs are likely not as informative for MACE for the dimer dataset.
MACE models trained on a new dataset acquired using the active learning procedure would likely perform better than those trained on this dataset acquired using the BPNNs, which we leave for future work.
We note that the water systems considered here are simple enough that the fixed symmetry functions of BPNNs provide a good description of the systems, while future systems of interest, such as reactive systems, will require more flexibility and scalability to a larger number of elements than BPNNs are readily capable of.

%\bibliography{ref} 
%\bibliographystyle{apsrev4-2}
%

\end{document}